\documentstyle[overcite,epsf]{aipproc}

\begin{document}

\title{{Effect of Optical Coating and Surface Treatments
on Mechanical Loss in Fused Silica}}
\author{Andri M. Gretarsson, Gregory M. Harry, Steven D. Penn, Peter R. Saulson,
John J. Schiller, and William J. Startin }
\address{\vspace{-10pt}Department of Physics, Syracuse University, Syracuse, NY 13244-1130, U.S.A.}
\maketitle
\vspace{-20pt}
\begin{abstract}
We report on the mechanical loss in fused silica samples with various surface
treatments and compare them with samples having an optical coating.
Mild surface treatments such as washing in detergent or acetone were not found
to affect the mechanical loss of flame-drawn fused silica fibers stored in air.
However, mechanical contact (with steel calipers) significantly increased
the loss. The application of a high-reflective optical coating of the type used
for the LIGO test masses was found to greatly increase the mechanical loss of
commercially polished fused silica microscope slides. We discuss the implications
for the noise budget of interferometers. \vspace{-2pt}
\end{abstract}

\section{Introduction}
\vspace{-4pt}
In samples made of high $Q$ materials, such as fused silica or sapphire, a
damaged or optically coated surface can be the dominant source of
mechanical loss and could limit our ability to reduce
thermal noise in interferometers.  We apply a general method
for quantifying surface loss to measurements of samples with
optical coatings and differing surface treatments.  This enables us to
estimate the effect of coatings on the internal mode thermal noise of
interferometer test masses as well as the effect of suspension
filament surface damage on the pendulum mode thermal noise.

\vspace{-5pt}
\section{Quantifying Surface Loss}
\vspace{-4pt}
Surface loss may be quantified by the
dissipation depth $d_s$, defined by\cite{Andri}
\vspace{-3pt}
\begin{equation}
\phi=\phi_{bulk}(1+\mu\frac{d_s}{V/S}),
\label{dissipation depth}
\end{equation}
\vspace{-15pt}\newline where $\phi=1/Q$ is the measured loss
angle of the sample when all sources of extrinsic loss (such as
recoil damping or clamping friction) have been eliminated,
$\phi_{bulk}$ is the loss angle of the bulk material, $V$ is the
volume of the sample, and $S$ is the surface area. The unitless
$\mu$ is a geometrical factor that takes into account the
relative amount of elastic deformation occurring at the surface
and hence the emphasis placed on the condition of the surface due
to the sample geometry and mode of oscillation. The geometrical
factor $\mu$ is of order unity for simple geometries so that, as
a rule of thumb, surface loss tends to dominate when $d_s$ is
greater than the volume to surface area ratio.  For fibers in
transverse oscillation $\mu=2$, while for ribbon or microscope
slide geometries in transverse oscillation $\mu=3$. Although
$\phi_{bulk}$ and $d_s$ may in general be functions of frequency,
no frequency dependence was seen in our measurements.  In this
paper we will use the constant value $\phi_{bulk}=3\times 10^{-8}$
for the loss angle of bulk fused silica. \vspace{-5pt}

\section{Surface Treatment of Uncoated Samples}
\begin{figure}
\hspace{0.6cm}
\epsfysize=8cm
\leavevmode
\epsfbox{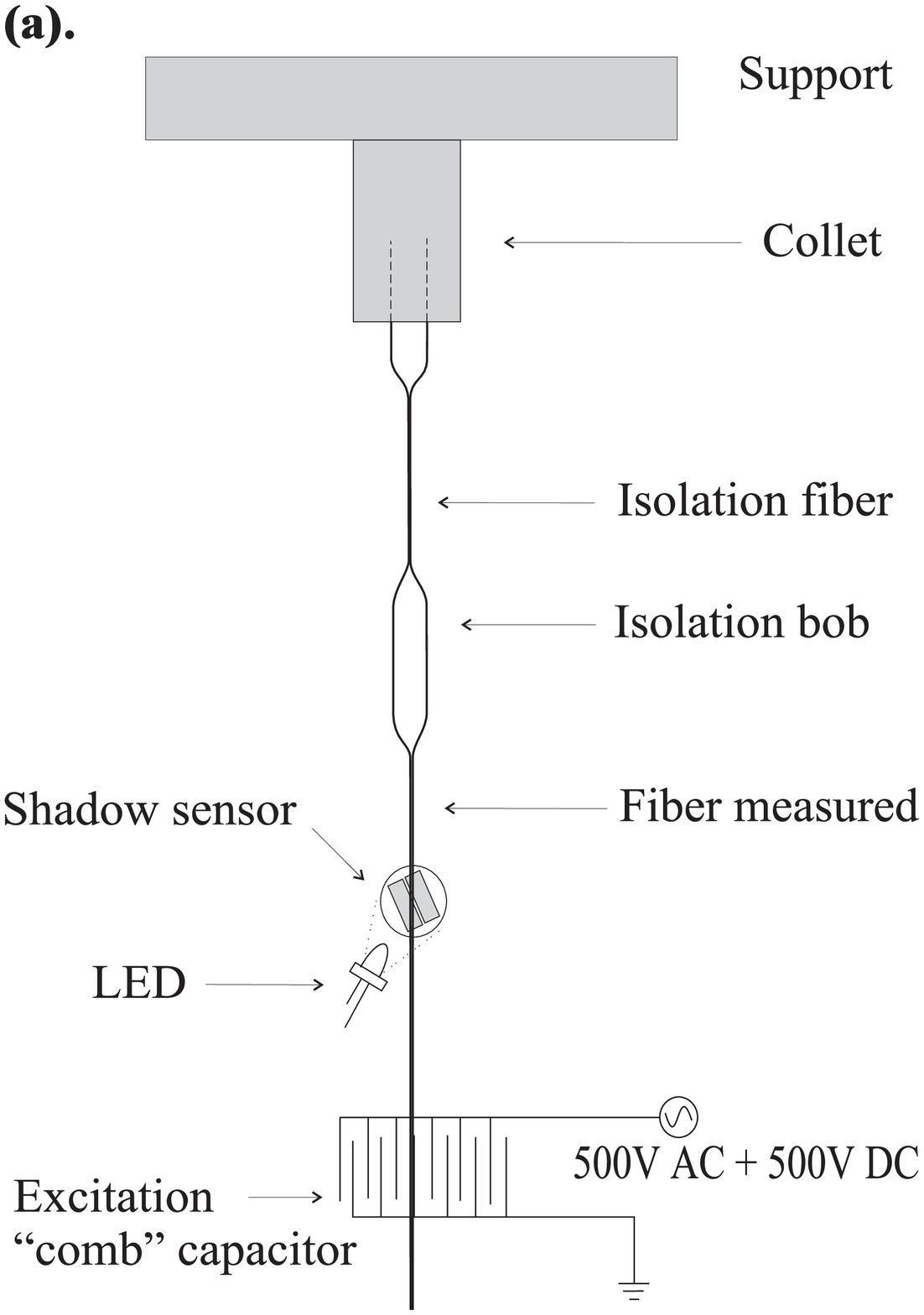}
\hspace{1.8cm}
\epsfysize=8cm
\leavevmode
\epsfbox{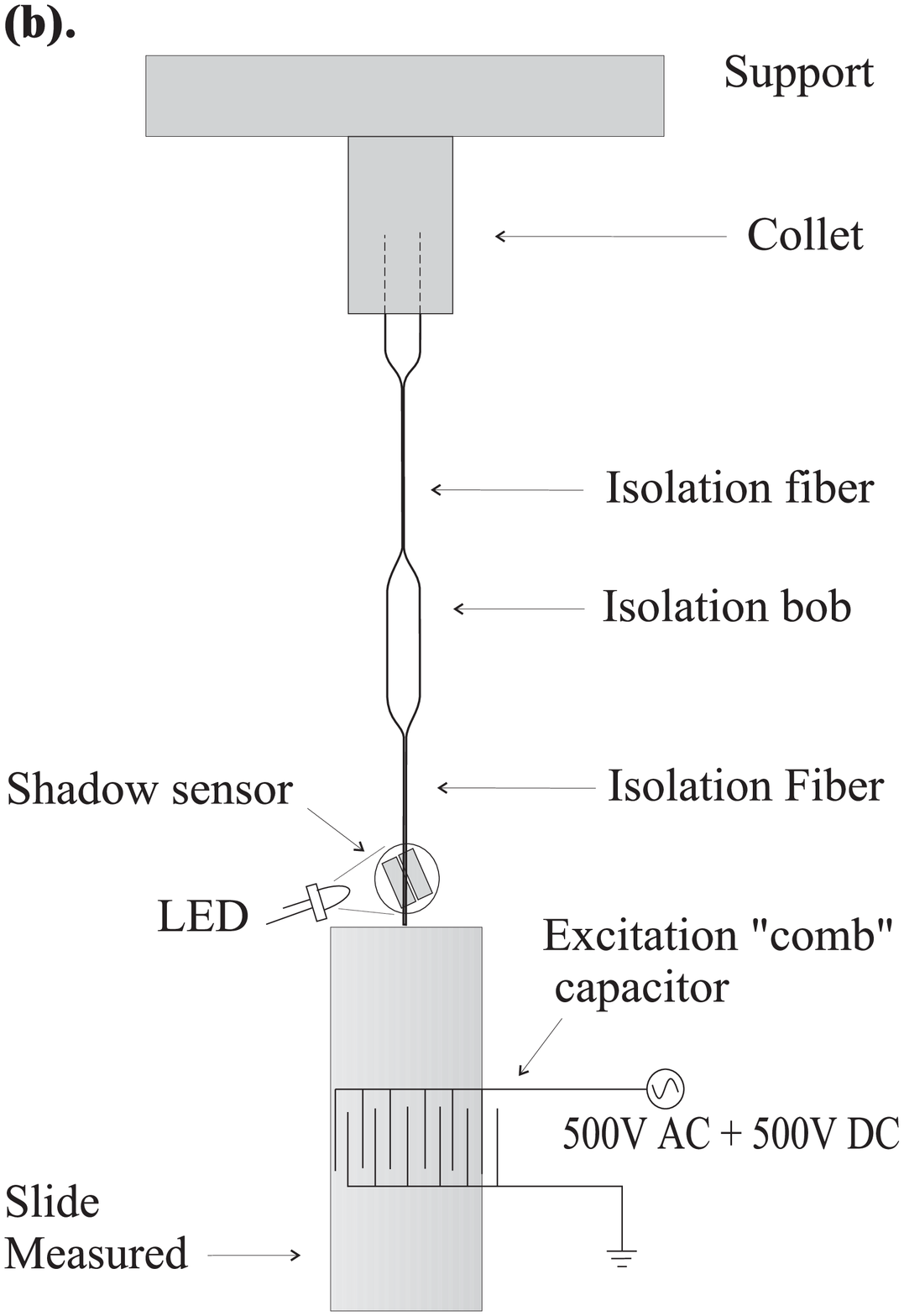}
\caption{Schematic diagram of the experimental setups.
{\bf (a).}~Setup for measuring fiber $Q$\/s.
{\bf (b).}~Setup for measuring slide $Q$\/s.
The isolation bobs and fibers prevent the measured sample $Q$ from being degraded by
rubbing in the clamp and from recoil in the lower $Q$ support structure. %The signal
%from the split photodiode shadow sensor is fed through a differential amplifier,
%a bandpass filter, a lock-in amplifier and a computer data acquisition system.
}
\label{apparatus}
\vspace{-11pt}
\end{figure}

\vspace{-5pt}
For uncoated samples the dissipation depth provides a quantitative measure of
the physical condition of the surface.
By measuring the quality factor $Q$ of samples before and after different types of
surface treatment, we calculated the dissipation depth associated with each treatment.
We measured the quality factors of untreated and treated fused silica (Suprasil~2)
fibers drawn in a natural gas and oxygen flame. We also measured the
quality factor of a fused silica (Suprasil~2) microscope slide, both
as supplied (mechanically polished) and as subsequently etched.
Using an apparatus specifically
designed for the purpose of reducing extrinsic sources of loss (Fig.~\ref{apparatus}a)
we were generally able to reduce extrinsic losses sufficiently so that the dominant sources
of loss remaining were thermoelastic loss, bulk loss, and surface loss.\cite{Andri}
In each case we measured the quality factors at frequencies where
thermoelastic loss was negligible.
In this regime the quality factors were frequency independent, although
in a minority of cases random mode-to-mode differences in $Q$ were apparent.
This was most likely due to residual sources of excess loss.
To reduce the systematic error due to such residual sources of excess loss we took the
highest~$Q$ mode to be indicative of the quality factor resulting from bulk loss and
surface loss alone.

To investigate the effects
of washing surfaces in solvents we wiped a fiber with paper wipes
(Kimwipes$^{\mathrm Tm}$) saturated with acetone.
We also agitated a fiber in an ultrasonic bath of detergent and warm
tap-water for a half hour, followed by a half hour ultrasonic bath of warm tap-water,
followed by a second rinse with a stream of distilled water.  After measuring the
Q we then waited 14 days with the fiber under vacuum ($\approx 10^{-6}$~Torr)
and re-measured the Q.
In an attempt to simulate the effects of hydroxy-catalysis bonding\cite{Guo}
(silicate bonding) of fused silica surfaces we washed a fiber with ethyl alcohol and then
submerged it in a 0.5 Molar solution of KOH and distilled water for 24.6~hrs,
then rinsed in distilled water. Also,
to investigate the effect of mechanical damage we lightly pinched
two fibers at 1~cm intervals with stainless steel measurement calipers.

To remove the outer surface entirely (and with it any mechanical surface damage)
we etched three fused silica fibers in solutions of hydrofluoric acid (HF) and
distilled water. After etching, the fibers were rinsed with distilled water.
The first etch was performed on a fiber of diameter $120\pm20~\micron$ and the
etch removed $1.5\pm0.5~\micron$ from the surface. The second etch was
performed on one of the fibers previously pinched with calipers. It had a pre-etch
diameter of $840\pm50~\micron$ and the etch depth was $45\pm3~\micron$.
The third etch was performed on a fiber of pre-etch diameter
$350\pm60~\micron$. The etch depth was $90\pm28~\micron$.
Finally, we etched the microscope slide.  As supplied, the microscope slide surface
had received a commercial 80-50 (scratch-dig) polish. The etch removed
$100~\micron$ from this surface.

Table~\ref{uncoated treatments} summarizes our results.
\begin{table}[h!]
\caption{Dissipation depth for different surface treatments.}
\label{uncoated treatments}
\begin{center}
\begin{tabular}{lllr}
sample  &treatment  &$d_s$ [$\mu$m] &$\Delta d_s$ [$\mu$m]\hspace{3pt}\tablenote{
Change in $d_s$ from the as drawn or as supplied state. The uncertainty in $\Delta d_s$
is not the root of the quadratic sum of uncertainties in $d_s$ since not all the variables involved in
calculating $d_s$ are independent between treatments.}\\    %&$\alpha_s$ [pm] \\
\hline
Fiber B &as drawn                       &$180\pm50$         &\\             %&$6.0\pm1.5$\\
Fiber B &acetone                        &$200\pm50$         &$17\pm10$\\    %&$6.5\pm1.6$\\
Fiber C &as drawn                       &$190\pm30$         &\\             %&$6.2\pm0.7$\\
Fiber C &calipers                       &$310\pm40$         &$124\pm20$\\   %&$10.3\pm1.0$\\
Fiber F &as drawn                       &$190\pm40$         &\\             %&$6.2\pm1.1$\\
Fiber F     &detergent solution, rinse  &$190\pm40$         &$-1\pm~9$\\    %&$6.2\pm1.1$\\
Fiber F &after 14 days in vacuum        &$160\pm30$         &$-25\pm10$\\   %&$5.4\pm0.9$\\
Fiber F &1.5~$\mu$m HF etch         &$220\pm50$         &$31\pm12$\\    %&$7.2\pm1.3$\\
Fiber I &as drawn                       &$340\pm50$         &\\             %&$11.3\pm1.1$\\
Fiber I &calipers                       &$620\pm100$        &$281\pm74$\\   %&$20.6\pm2.6$\\
Fiber I &45$\mu$m HF etch               &$100\pm20$         &$-244\pm30$\\  %&$3.3\pm0.5$\\
Fiber L &as drawn                       &$210\pm50$         &\\             %&$7.1\pm1.7$\\
Fiber L &KOH, 0.5 M solution, rinse&$310\pm80$          &$95\pm29$\\    %&$10.2\pm2.3$\\
Fiber M &as drawn                       &$100\pm20$         &\\             %&$3.2\pm0.7$\\
Fiber M &90~$\mu$m HF etch          &$180\pm40$         &$86\pm25$\\    %&$6.0\pm1.2$\\
Slide C &as supplied (polished)     &$860\pm140$        &\\             %&$28\pm4$\\
Slice C     &100~$\mu$m HF etch         &$850\pm160$        &$10\pm140$     %&$28\pm4$
\end{tabular}
\end{center}
\vspace{-10pt}
\end{table}
For three of the fibers, with surfaces as drawn, the dissipation depth
is around $200~\micron$.  Fiber I has a surface that is initially worse
(higher $d_s$) while Fiber M has a surface that is initially better.
Although an effort was made not to touch the
surfaces of the fibers with fingers or other objects during handling,
conditions were not stringently uniform.  The fibers were also
stored for varying durations in clean glass tubes and could come into light contact
with the inner surface of the tubes.  Depending on the storage time or
amount of contact, some deviation in $d_s$ can be expected.

\begin{figure}
\begin{center}
\epsfysize=6.5cm
\leavevmode
\epsfbox{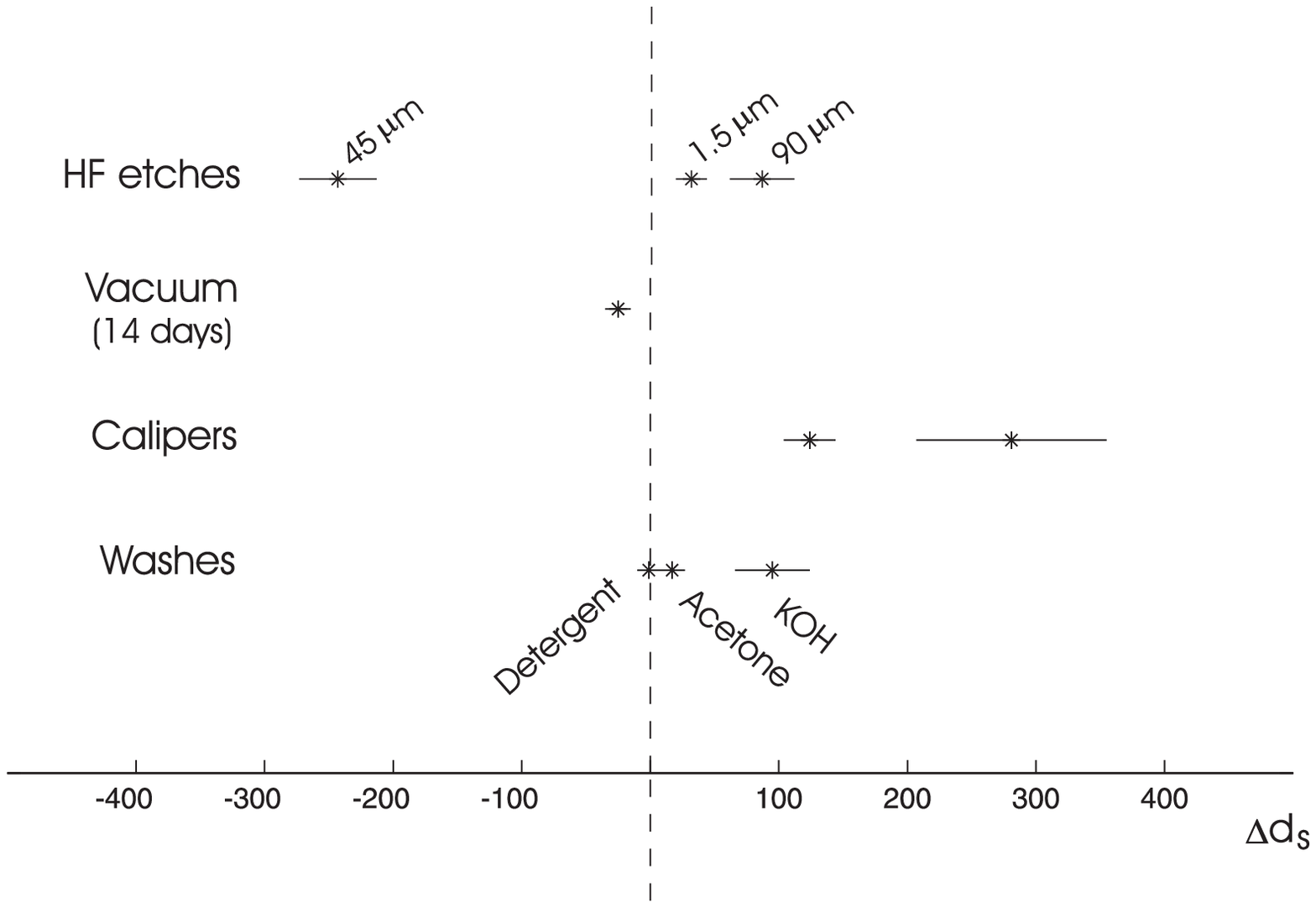}
\end{center}
\caption{Change in the condition of fiber surfaces as compared to the initial surface as drawn,
$\Delta d_s$ vs treatment. Horizontal lines mark the uncertainty in $\Delta d_s$.}
\label{delta_ds}
\vspace{-10pt}
\end{figure}

Figure~\ref{delta_ds} shows how strongly different treatments affected
the surface of fibers.
Most of the treatments either produced no change in the condition
of the surface or they made it only slightly worse.  However, pinching the
surface at regular (1~cm) intervals with calipers significantly increased
the dissipation depth, possibly due to small cracks formed by the mechanical
contact. Similarly, the mechanically polished microscope slide had the
highest measured $d_s$. It is interesting to note that the
surface of Fiber I, after being significantly damaged by calipers,
was restored to a condition better than as drawn (or perhaps
more appropriately better than ``as stored and handled'')
by the $45~\micron$ HF etch.  The resulting dissipation depth agrees with the best
as drawn case, having a value of about $100~\micron$.

The question arises why most samples undergoing HF etches did not show
significant surface improvement.  In the case of the severely damaged slide,
we believe the etch was too shallow.  After etching, hairline scratches
on the slide were visible to the naked eye.  Etching opens up microscopic cracks
imparted by the polishing process and their presence, post-etch, is evidence that
the surface was still damaged.  As for the fiber etches, only one (the etch of Fiber~I)
resulted in an improved dissipation depth.  This may be due to the fact that Fiber~I
had an as drawn dissipation depth somewhat higher than any other fiber and may
have been inadvertently damaged between drawing and installation in
our apparatus.  Mechanical damage can be repaired by HF etching~\cite{kreidl,doremus}
(though the etches must be sufficiently deep).  The HF etch may thus have removed the
damaged surface of Fiber I, reducing the dissipation depth from the initially measured
value.

We should not neglect the possibility that chemical contamination
of the surface, in particular contamination with atmospheric water,\cite{doremus}
may also lead to increased loss.
The ubiquity of $d_s$ values in the range 100-200~$\micron$ could
be due to the difficulty of isolating samples from atmospheric water.
This would also explain the failure of the etches to reduce the dissipation depth
below this range.

\section{Surface Loss due to Optical Coating}
\vspace{-5pt}
The surface loss due to optical coatings was investigated by measuring
the quality factors of the modes of fused silica slides.  We measured the $Q$\/s
for three slides of dimensions 76~mm~$\times$ 25~mm~$\times$~1~mm.
The slides were suspended below a monolithic, fused-silica
isolation system, as shown in Fig.~\ref{apparatus}b.  The slides' vibration
was monitored by positioning the LED and shadow sensor around the suspending
fiber directly above the slide.

Two of the three slides, A and B, were optically coated while the third
slide, C, was retained uncoated as a control.  (Slide C was later etched as
reported in the preceding section.)  As supplied, the slides had received a
commercial 80-50 polish.
The optical coating applied to Slide A and Slide B was a
high-reflective (HR) coating $2.4~\micron$ thick
consisting of 14 layers of alternating SiO$_2$ and Ta$_2$O$_5$.  The
slides were coated by ion beam sputtering in the same coating run as optics
for LIGO by Research Electro-Optics Corporation in Boulder Colorado.  After the
coating run they were baked at 450$^\circ$C to relieve stress.

The quality factors for each measured resonant mode
of slides A and B and equivalent dissipation depths
are shown in Table~\ref{coated}.  The quality factors and equivalent
dissipation depths for the measured modes of the uncoated
Slide C (as supplied) are given for comparison.
\begin{table}[h!]
\begin{center}
\caption{Resonant $Q$\/s  and equivalent dissipation depth in
coated slides.}
\label{coated}
\begin{tabular}{clcrcr}
Slide   &Surface treatment  &Mode    &Frequency &$Q$                        &$d_s$~[$\mu$m]\hspace{0.6cm}  \\
\hline \\
A   &HR-coating with no             &2      &1022 Hz        &$1.1\pm 0.5\times 10^5$   &$46\pm 21   \times 10^3$\\
    &visible damage             &3      &1944   Hz      &$1.6\pm 0.1\times 10^5$   &$32\pm 3    \times 10^3$\\
    &                                   &4      &2815   Hz      &$1.6\pm 0.1\times 10^5$   &$32\pm 3    \times 10^3$\\
B   &HR-coating damaged         &2      &952    Hz      &$3.1\pm 0.2\times 10^4$   &$160 \pm 15\times 10^3$\\
    &at top by flame.               &3      &1851   Hz      &$1.6\pm 0.1\times 10^5$   &$32\pm 3    \times 10^3$\\
B   &Damaged region removed     &2      &962    Hz      &$1.3\pm 0.1\times 10^5$    &$39\pm4    \times 10^3$\\
C   &Uncoated, as supplied      &2      &1188   Hz      &$4.0\pm 0.2\times 10^6$    &$1.1\pm 0.2\times 10^3$\\
    &(``80-50'' polish)         &3  &2271   Hz      &$4.9\pm 0.3\times 10^6$   &$0.86\pm0.14\times 10^3$\\
\end{tabular}
\end{center}
\vspace{-10pt}
\end{table}

The coated Slide B was suspended from the center of one of its short edges, as
shown in Fig.~\ref{apparatus}b. When the supporting fiber was connected to the slide
using a hydrogen-oxygen torch the coating became visibly damaged.
Where the flame from the torch contacted the coating, the coating took
on a milky appearance.  This occurred in a crescent shape approximately
3~mm across at the top of the slide.  The high value
of $d_s$ for the second mode is believed to be due to this
damage.  To test this, the top 5~mm of this slide were immersed in a
50\% solution (by weight) of HF and water for about 6 hours.
Rinses with distilled water were applied periodically to remove
flakes of the coating. The etch removed most of the damaged
part of the coating and the  $Q$  was re-measured.  The  $Q$  and
dissipation depth of the second mode was now of the same magnitude as that
measured for the third mode and for all modes of Slide B.

The coated Slide A was hung from a corner rather than from the
center of the top edge. This was because, in the corner, the
fused silica substrate was masked (by the supports) during the
coating process.  This left a region with no optical coating
about 1~mm in radius and centered on the corner. The fiber was
very carefully welded to the slide at this point. While some heat
from the torch certainly reached the coated region, no damage to
the coating could be seen afterwards. Both  modes of Slide A
showed similar Qs and similar dissipation depths as the modes of
Slide B after the damaged region was removed.  Since the uncoated
Slide C has significantly less dissipation than the coated
slides, and since the coated slides all show approximately the
same level of dissipation, we conclude that the high dissipation
depth associated with the coated slides, $d_s\approx3$~cm, is a
result of the HR optical coating.  If the coating is modeled as
having homogeneous loss $\phi_{coat}$, and we assume that loss in
the interface between the coating and substrate is negligible,
then $\phi_{coat}$ is simply~\cite{Andri}
\begin{equation}
\phi_{coat}=\frac{d_s}{h}\phi_{bulk}
\end{equation}
where $h$ is the thickness of the coating.  Thus, we obtain the
preliminary result,
\begin{equation}
\phi_{coat} \approx 4\times10^{-4}.
\end{equation}

If our measurements are characteristic of the coatings for LIGO,
this would lead to noticeably increased thermal noise for the LIGO
test masses. However, the surfaces of the slides did not receive
the same treatment prior to the coating as the LIGO test masses.
They were not superpolished and no particular efforts were made
to ensure the absolute cleanliness of the surfaces. It is
possible that the interface between the coating and the silica is
more lossy than a polished surface interface would be.
Superpolished samples of fused silica have been obtained and
research is continuing to determine the loss in superpolished and
coated samples.

\section{Implications for Thermal Noise}
\vspace{-1pt}
Surface loss in the filaments suspending LIGO test masses
could have implications for the interferometer noise budget.\cite{ribbon_paper}
Surface loss associated with fibers implies a lower limit
on the level of pendulum mode thermal noise achievable using thin ribbon suspensions.
While dissipation dilution implies reduced pendulum mode thermal noise as the ribbon
thickness is reduced, the effects of surface loss are increased. The result is a
diameter-independent lower limit for the pendulum mode thermal displacement noise
spectral density
\begin{equation}
x^2_{\mathrm{min}}(\omega) =
\frac{24k_BTg}{ML^2\omega^5}\,\sqrt{\frac{Y}{12\sigma}}\,d_s\phi_{\mathrm{bulk}},
\label{x_min}
\end{equation}
where $\omega$ is the angular frequency, $k_B$ is Boltzmann's constant, $T$ is
the temperature, $g$ is the acceleration due to gravity, $M$ is the suspended mass,
$L$ is the length of the suspension, $Y$ is Young's modulus, and $\sigma$ is the
stress in the suspending ribbons.  For typical values of the parameters and
$d_s=200~\mu$m, we have
$$x_{\mathrm min}(\omega=2\pi\times10~{\mathrm Hz})\approx6\times10^{-20}~\mathrm{ m/\sqrt{Hz}}.$$
While this is sufficient for the goals
of LIGO II, it is clear from the dependence on $d_s$ that
mechanical surface damage such as is induced by calipers must be prevented.

Surface loss due to optical coatings may significantly increase the thermal noise
due to internal modes of the test masses.
To relate the dissipation depth measured for an optical coating to the internal mode
thermal noise we follow the work of Levin\cite{Levin} and
Bondu~{\it et al}\cite{Bondu}.  This enables an approximate calculation of the
relevant~$\mu$. Using Eq.~\ref{dissipation depth} we obtain after some analysis a
preliminary estimate for the test-mass loss angle,
%\vspace{-8pt}
\begin{equation}
\phi\approx \phi_{bulk}(1+0.9\frac{d_s}{w}),
\end{equation}
%\vspace{-8pt}\newline
where $w$ is the radius at which the amplitude of the beam falls
to $1/e$ of its maximum value. Since $w\approx3.5$~cm in LIGO I,
and will be of the same order in LIGO II, it is clear that if
$d_s\approx3$~cm, as measured for the coated slides, then the HR
coating will be a significant contributor to test mass thermal
noise.

\end{document}